\documentclass[reprint,twocolumn,superscriptaddress,amsmath,amssymb,showpacs,prl]{revtex4-1}
\usepackage{hyperref}
\pdfoutput=1 

\usepackage{dcolumn}
\usepackage{epsfig}
\usepackage{graphics}
\usepackage[latin1]{inputenc} 
\usepackage[T1]{fontenc}
\usepackage{bm}
\usepackage{hyperref}

\begin{document}

\title{Is Cement a Glassy Material?}

 \author{M. Bauchy}
 \email[Contact: ]{bauchy@ucla.edu}
 \homepage[\\Homepage: ]{http://mathieu.bauchy.com}
 \affiliation{Physics of AmoRphous and Inorganic Solids Laboratory (PARISlab), Department of Civil and Environmental Engineering, University of California, Los Angeles, CA 90095, United States}
 \author{M. J. Abdolhosseini Qomi}
 \affiliation{Concrete Sustainability Hub, Department of Civil and Environmental Engineering, Massachusetts Institute of Technology, 77 Massachusetts Avenue, Cambridge, MA 02139, United States}
 \author{Franz-Joseph Ulm}
 \affiliation{Concrete Sustainability Hub, Department of Civil and Environmental Engineering, Massachusetts Institute of Technology, 77 Massachusetts Avenue, Cambridge, MA 02139, United States}
 \affiliation{MIT-CNRS joint laboratory at Massachusetts Institute of Technology, 77 Massachusetts Avenue, Cambridge, MA 02139, United States}
 \author{Roland Pellenq}
 \affiliation{Concrete Sustainability Hub, Department of Civil and Environmental Engineering, Massachusetts Institute of Technology, 77 Massachusetts Avenue, Cambridge, MA 02139, United States}
 \affiliation{MIT-CNRS joint laboratory at Massachusetts Institute of Technology, 77 Massachusetts Avenue, Cambridge, MA 02139, United States}
 \affiliation{Centre Interdisciplinaire des Nanosciences de Marseille, CNRS and Aix-Marseille University, Campus de Luminy, Marseille, 13288 Cedex 09, France}

\date{\today}


\begin{abstract}
The nature of Calcium--Silicate--Hydrate (C--S--H), the binding phase of cement, remains a controversial question. In particular, contrary to the former crystalline model, it was recently proposed that its nanoscale structure was actually amorphous. To elucidate this issue, we analyzed the structure of a realistic simulation of C--S--H, and compared the latter to crystalline tobermorite, a natural analogue to cement, and to an artificial ideal glass. Results clearly support that C--S--H is amorphous. However, its structure shows an intermediate degree of order, retaining some characteristics of the crystal while acquiring an overall glass-like disorder. Thanks to a detailed quantification of order and disorder, we show that its amorphous state mainly arises from its hydration.
\end{abstract}

\maketitle

\section{Introduction}
\label{sec:intro}


Calcium-silicate-hydrate (C--S--H) is the main hydration product in Portland cement and acts as a binding phase in the paste \cite{taylor_cement_1997} . Thus, it is responsible for its strength, durability and creep properties \cite{vandamme_nanogranular_2010}.

Despite its ubiquitous presence in the built environment, the structure of C--S--H at the nanoscale remains controversial. As X-Ray diffraction patterns from C--S--H have shown to exhibit only a few broad and weak diffraction maxima, it has been described as an amorphous material \cite{gartner_proposed_2000,cong_ca_1997,mandaliev_exafs_2010}. However, most experimental studies \cite{cong_29si_1996,allen_analysis_2007,groves_tem_1986,grangeon_nature_2013,grangeon_x-ray_2013} suggest that its structure is close to the one of tobermorite, although their composition differ. Hence, it is still unclear whether C--S--H should be considered as a crystalline or an amorphous material. Fortunately, a realistic atomistic model of C--S--H has recently been reported \cite{pellenq_engineering_2008,skinner_nanostructure_2010,abdolhosseini_qomi_combinatorial_2014,abdolhosseini_qomi_applying_2013}, thus opening the way to elucidate its nature.

To quantify the atomic order in C--S--H, we analyzed the previously mentioned atomistic model. We compared its structure with the one of tobermorite, its crystal analogue, while taking care of rescaling the results to account for the difference of chemistry. The structure was also compared to the one of an ideal artificial glass, formed by quickly heating and cooling a C--S--H system. By doing a consistent analysis of the structure of those three systems, simulated in the same conditions and with the same potential, we were able to assess their relative atomic order both at the short and medium range order. 

The article is organized as follows. In section \ref{sec:met}, we present the numerical model and the methodology that we used to simulate C--S--H and the corresponding crystal and glass. Then, we analyze the atomic structure of C--S--H as compared to the ones of the crystal and the glass in the short range oder in \ref{sec:sro}. This comparison is then extended to the medium range order in \ref{sec:mro}.  Concluding remarks are presented in secion \ref{sec:conclu}.

\section{Methods}
\label{sec:met}

In this section, we detail the model and the procedure used to simulate C--S--H and its equivalent crystal and glass.

\subsection{A realistic model of C--S--H}

To describe the disordered molecular structure of C--S--H, Pellenq et al. \cite{pellenq_realistic_2009} proposed a realistic model for C--S--H with the stoichiometry of (CaO)$_{1.65}$(SiO$_2$)(H$_2$O)$_{1.73}$. We generated the C--S--H model by introducing defects in an 11\AA\ tobermorite \cite{hamid_crystal-structure_1981} configuration, following a combinatorial procedure. 11\AA\ tobermorite consists of pseudo-octahedral calcium oxide sheets, which are surrounded by silicate tetrahedral chains. The latter consists of bridging oxygen atoms and $Q^2$ silicon atoms (having two bridging and two non-bridging terminal oxygen atoms) \cite{abdolhosseini_qomi_evidence_2012}. Those negatively charged calcium-silicate sheets are separated from each other by an interlayer spacing, which contains interlayer water molecules and charge-balancing calcium cations. While the Ca/Si ratio in 11\AA\ tobermorite is 1, this ratio is increased to 1.65 in the present C--S--H model through randomly removing SiO$_2$ groups. The defects in silicate chains provide possible sites for adsorption of extra water molecules. The adsorption of water molecules in the structurally defected tobermorite model was performed via the Grand Canonical Monte-Carlo method, ensuring equilibrium with bulk water at constant volume and room temperature. The REAXFF potential \cite{manzano_confined_2012}, a reactive potential, was then used to account for the reaction of the interlayer water with the defective calcium-silicate sheets. The use of the reactive potential allows observing the dissociation of water molecules into hydroxyl groups. More details on the preparation of the model and on the multiple validations with respect to experiments can be found elsewhere \cite{pellenq_engineering_2008,skinner_nanostructure_2010,abdolhosseini_qomi_combinatorial_2014,abdolhosseini_qomi_applying_2013}.

\subsection{Simulation of C--S--H}

We simulated the C--S--H model previously presented, made of 501 atoms, by molecular dynamics using the LAMMPS package \cite{plimpton_fast_1995}. To this end, we used the REAXFF potential \cite{manzano_confined_2012} with a time step of 0.25fs. We first relaxed the system at zero pressure and 300K during 2.5ns in the NPT ensemble and made sure that the convergence of the energy and volume was achieved. We then run a 25ps simulation in the NVT ensemble for statistical averaging. 

\subsection{Simulation of the crystal analogue}

We chose to study 11\AA\ tobermorite as it is considered as a natural crystal analogue for C--S--H (see section \ref{sec:intro}). This choice was also motivated by the fact that the simulated C--S--H sample was prepared by introducing defects inside tobermorite (see section \ref{sec:met}).

For consistent comparison with C--S--H, this system was simulated using the REAXFF potential, i.e. the same as the one we used for C--S--H, the same time step and following the same procedure. We started from an initial 11\AA\ tobermorite cell \cite{hamid_crystal-structure_1981} composed of 288 atoms, relaxed it at zero pressure and 300K during 2.5ns in the NPT ensemble, checked the convergence of the volume and the energy and run a 25ps simulation in the NVT ensemble for statistical average. Note that the use of the REAXFF potential does not induce any significant change of volume nor structural modification with respect to the starting configuration \cite{hamid_crystal-structure_1981}. 

It should be noted that 11\AA\ tobermorite does not have the same composition nor the same density as C--S--H. To be able to compare the results with C--S--H, we rescaled all the computed properties to take this difference into account. In practice, this is achieved by replacing the concentration of every species by the ones in C--S--H. In the following, we will refer to the results obtained with the tobermorite system and after the mentioned rescaling as being the properties of the \textit{crystal}.

\subsection{Simulation of the glass analogue}

We aimed at comparing the structure of C--S--H with the one of an ideal glass. To that end, we created an artificial glass analogue of C--S--H by heating and cooling a C--S--H configuration. Note that this methodology is commonly used to simulate glassy materials; e.g. one usually prepare glassy silica by heating and cooling quartz \cite{bauchy_angular_2011}.

We started from the relaxed configuration of C--S--H. The system was then instantly heated at 3000K and relaxed during 2.5ns at constant pressure in the NPT ensemble at this temperature, which is well above the melting temperature calcium silicate systems (for example 1813K for the dicalcium silicate crystal). This allowed the system to loose the memory of its initial configuration. We checked this by computing the root-mean-square displacement of each species and making sure that, at the end of the simulation, they were far larger than the size of the simulation box. The system was then gradually cooled from 3000K down to 300K in the NPT ensemble, with a cooling rate $q$ of 20, 40 and 80K/ps. During the melting and cooling phases, we imposed a pressure of 0.1GPa to prevent water molecules to leave the system at high temperature, which would lead to artificially large simulation box. However, this pressure is low with respect to the large pressure fluctuations in the system due to its small size (typically around 1GPa). Therefore, we do not expect this pressure to have any significant impact on the structure of the simulated glass. Once at 300K, the obtained glass was relaxed in the NPT ensemble at zero pressure during 2.5ns. At this stage, we checked the convergence of the energy and volume of the system. Eventually, we run a 25ps simulation in the NVT ensemble at 300K for statistical averaging. 

Note that the volume of the simulation box increases by 80\% while heated at 3000K. However, after being cooled, the volume of the system goes back to a value fairly close to the volume of the initial C--S--H box (larger by 3\%). Once again, to allow for a consistent comparison with C--S--H, we used the same REAXFF potential and the same time step during the entire procedure. In the following, we will refer to the results obtained with this quenched system as being the properties of the \textit{glass}.

\section{Short-range order}
\label{sec:sro}



\subsection{Total pair distribution function}

\begin{figure}
\centerline{
\includegraphics[height=8cm, angle=-90]{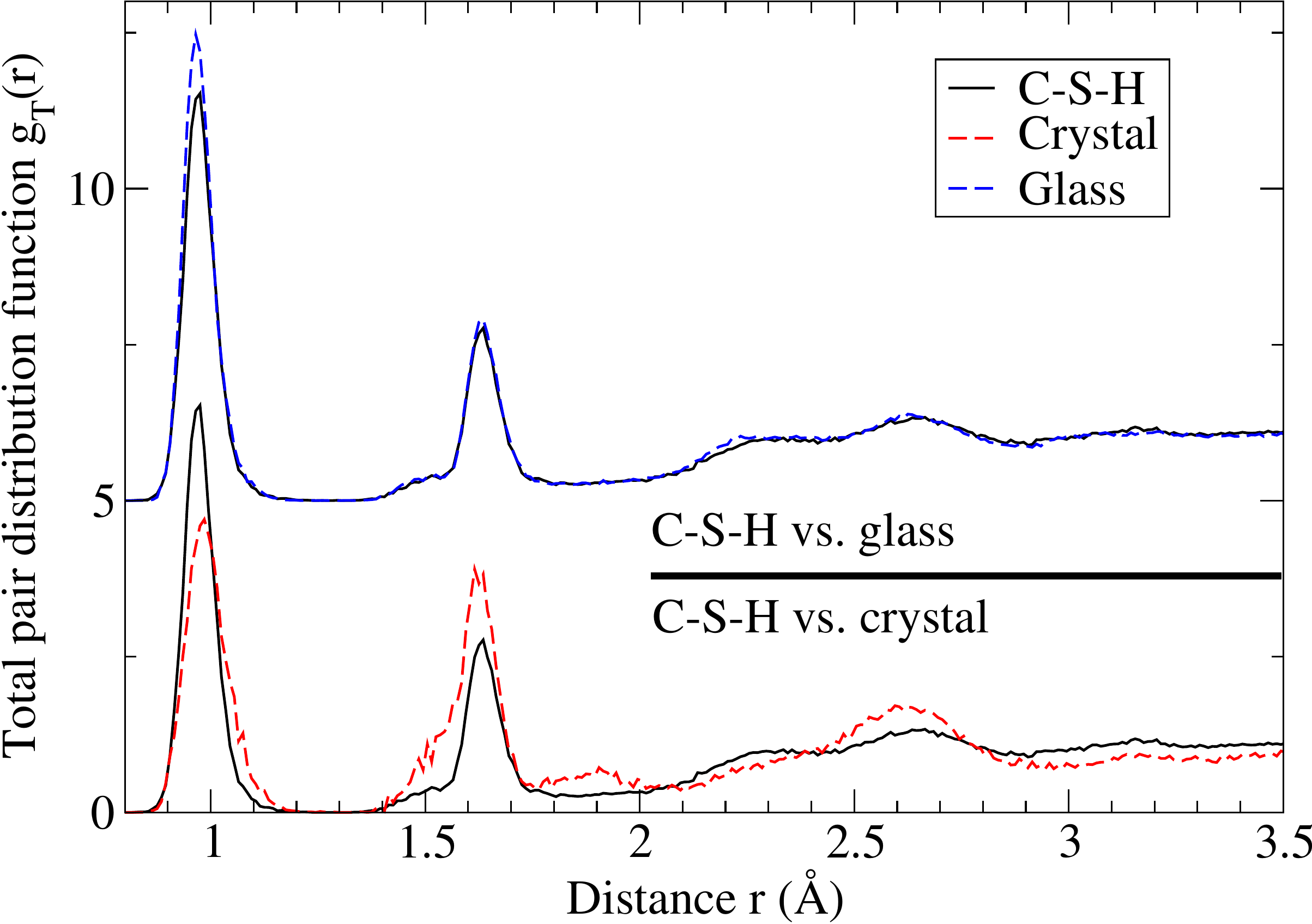}
}
\caption{Total pair distribution function of C--S--H, compared with the one of the corresponding glass (top) and of the crystal (bottom).}
\label{fig:gt}
\end{figure}

As a first step, we computed the total pair distribution function (PDF) $g_{\text{T}}(r)$ of the three systems from the knowledge of the partial PDF $g_{ij}(r)$:
\begin{eqnarray}
\label{eq:gt}
g_{\text{T}}(r) = \sum_{i,j=1}^n c_ic_j g_{ij}(r)
\end{eqnarray} where $c_i$ is the fraction of $i$ atoms (Si, Ca, H or O).

Fig. \ref{fig:gt} shows the total PDFs of C--S--H, compared with the one of the crystal and the glass. We note that the PDFs of those three systems does not show significant differences that would clearly allow distinguishing a glassy from a crystalline phase. They actually present a similar shape than the PDFs of silicate \cite{bauchy_structural_2012} and chalcogenide glasses \cite{micoulaut_structure_2013}. The first peak in the 1\AA\ region is associated to H-O bonds. The second peak in the 1.6\AA\ region corresponds to a superposition of the H-H and Si-O correlations. Following peaks are not easy distinguishable as they result of the superposition of several contributions of different pairs of atoms.

Overall, we observe a closer agreement of the PDF of C--S--H with the one of the glass than with the one of the crystal. This tends to show that C--S--H is mostly amorphous.

\subsection{Partial pair distribution function}

\begin{figure*}
\centerline{\includegraphics[height=15cm, angle=-90]{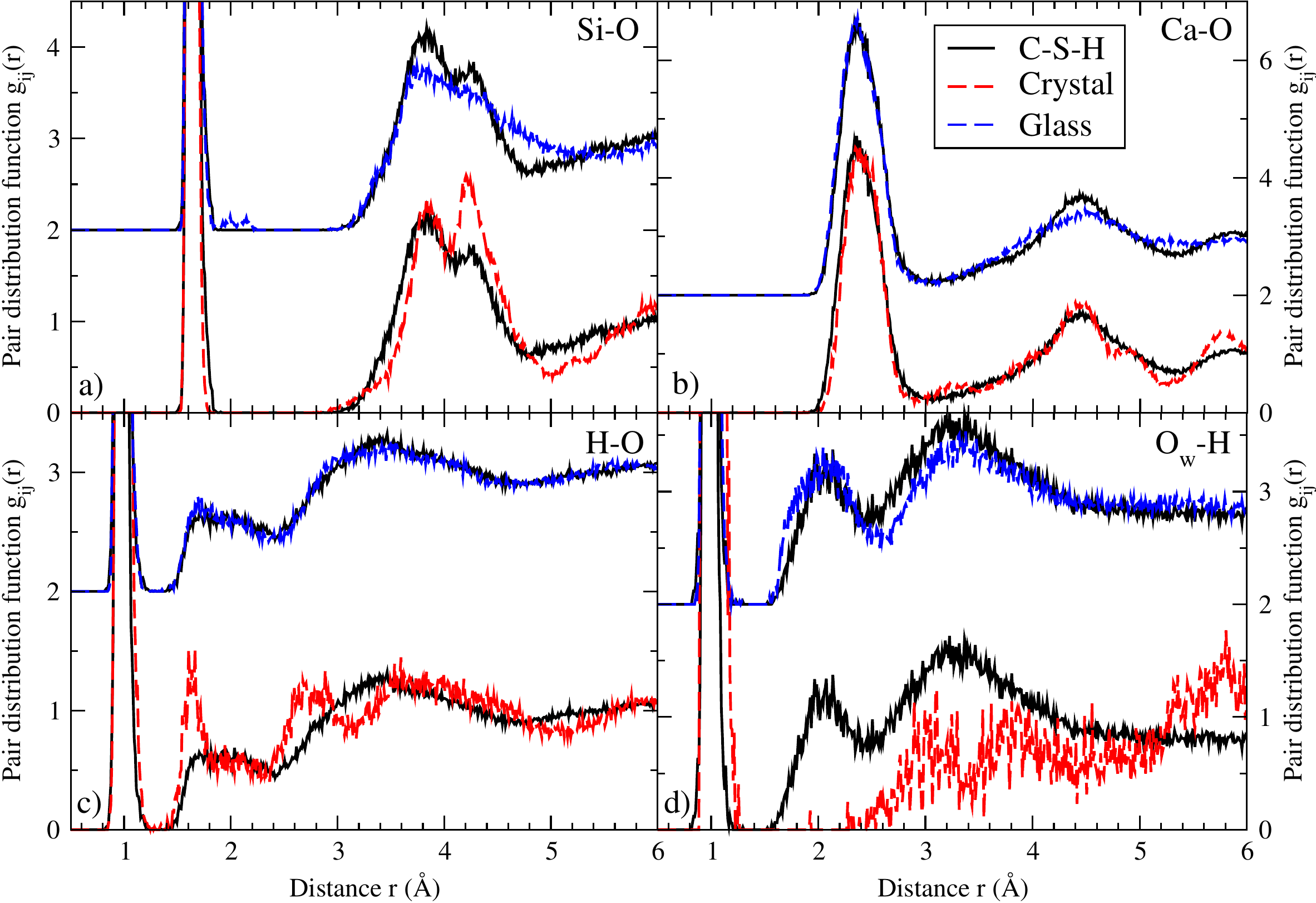}}
\caption{Partial distribution functions for the Si-O (a), Ca-O (b), H-O (c), and O$_{\text{w}}$-H pairs, where O$_{\text{w}}$ refers to oxygen atoms in water molecules. For each pair, the partial distribution function of C--S--H is compared with the one of the corresponding glass (top) and of the crystal (bottom).}
\label{fig:gij}
\centerline{\includegraphics[height=15cm, angle=-90]{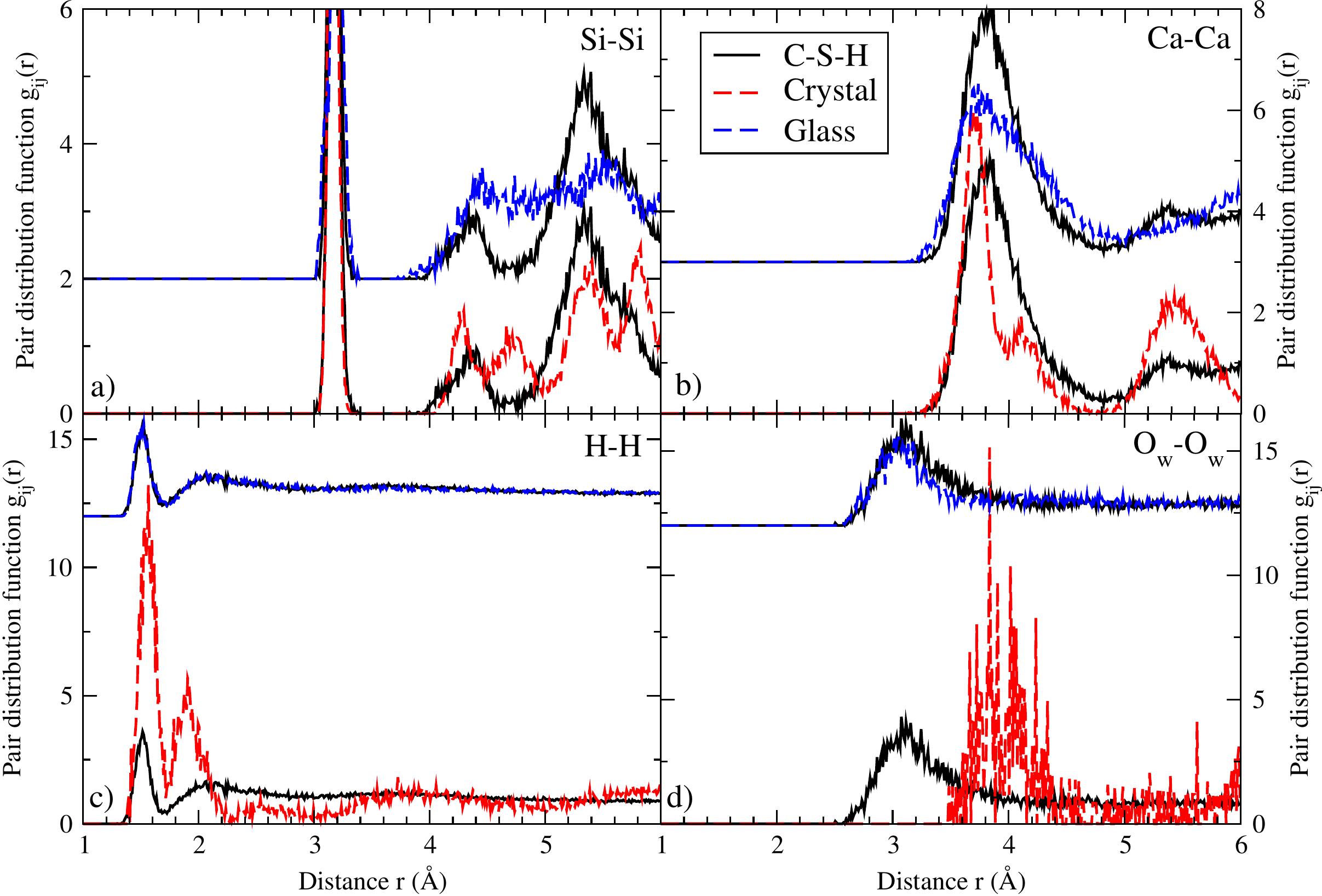}}
\caption{Partial distribution functions for the Si-Si (a), Ca-Ca (b), H-H (c), and O$_{\text{w}}$-O$_{\text{w}}$ pairs, where O$_{\text{w}}$ refers to oxygen atoms in water molecules. For each pair, the partial distribution function of C--S--H is compared with the one of the corresponding glass (top) and of the crystal (bottom).}
\label{fig:gij2}
\end{figure*}

To better characterize the order and disorder in C--S--H, we computed all the partial PDFs. Selected partial PDFs are shown in Fig. \ref{fig:gij}. Note that O$_w$ refers to oxygen atoms belonging to water molecules. Fig. \ref{fig:gij}a and \ref{fig:gij}b show that the short-range order of Si and Ca atoms is roughly similar in the crystal, in C--S--H and in the glass. In particular, the position and integral of the first peak do not show any significant change, which means that the atomic bond distance Si-O and Ca-O as well as the coordination number of the cations are the same. This is not surprising as glasses typically tend to retain the same local order as observed in the corresponding crystal.

However, some differences start to appear at larger scale. The second coordination shell peak of the Si-O partial PDF of C--S--H shows a bimodal distribution which is reminiscent of the one observed in the crystal. On the contrary, the glass only shows a smooth broad peak corresponding to the second coordination shell. This means that a certain degree of order is maintained in the silicate layers in C--S--H. In the case of the Ca-O partial, the PDF of the crystal shows a small peak around 3.2\AA\, which disappears in C--S--H. However, the peaks observed in the PDF of the glass are broader than the ones of C--S--H. Overall, C--S--H seems to present an atomic order that is intermediate between the ones of the crystal and the glass in the Si and Ca local environment.

On the contrary, the environment of hydroxyl groups (Fig. \ref{fig:gij}c) and water molecules (\ref{fig:gij}d) show a different behavior. Once again, the first peak is fairly similar in C--S--H, in the crystal  and in he glass, corresponding to similar bond distances and coordination numbers. However, structural correlations at larger scale ($r$>1.5 \AA) appear to be almost identical in C--S--H in the glass, and significantly differ from the ones observed in the crystal. This is a clear evidence the disorder observed inside C--S--H mainly arises from water molecules and hydroxyl groups, i.e. from the hydration of the network.

Since the structural differences between C--S--H, the glass and the crystal seem to appear at scales larger than the one of the first coordination shell, we analyzed second neighbor cation-cation and water-water correlations. Overall, the first peaks of the Si-Si (Fig. \ref{fig:gij2}a) and Ca-Ca (Fig. \ref{fig:gij2}b) partial PDFs of C--S--H are systematically broader than in the crystal, but sharper than in the glass. This confirms the fact that the calcium and silicate layers in C--S--H retain some crystal-like order, as opposed to H-H (Fig. \ref{fig:gij2}c) and water-water (Fig. \ref{fig:gij2}d) correlations which are essentially amorphous.

\section{Medium-range order}
\label{sec:mro}

\begin{figure}
\centerline{
\includegraphics[height=8cm, angle=-90]{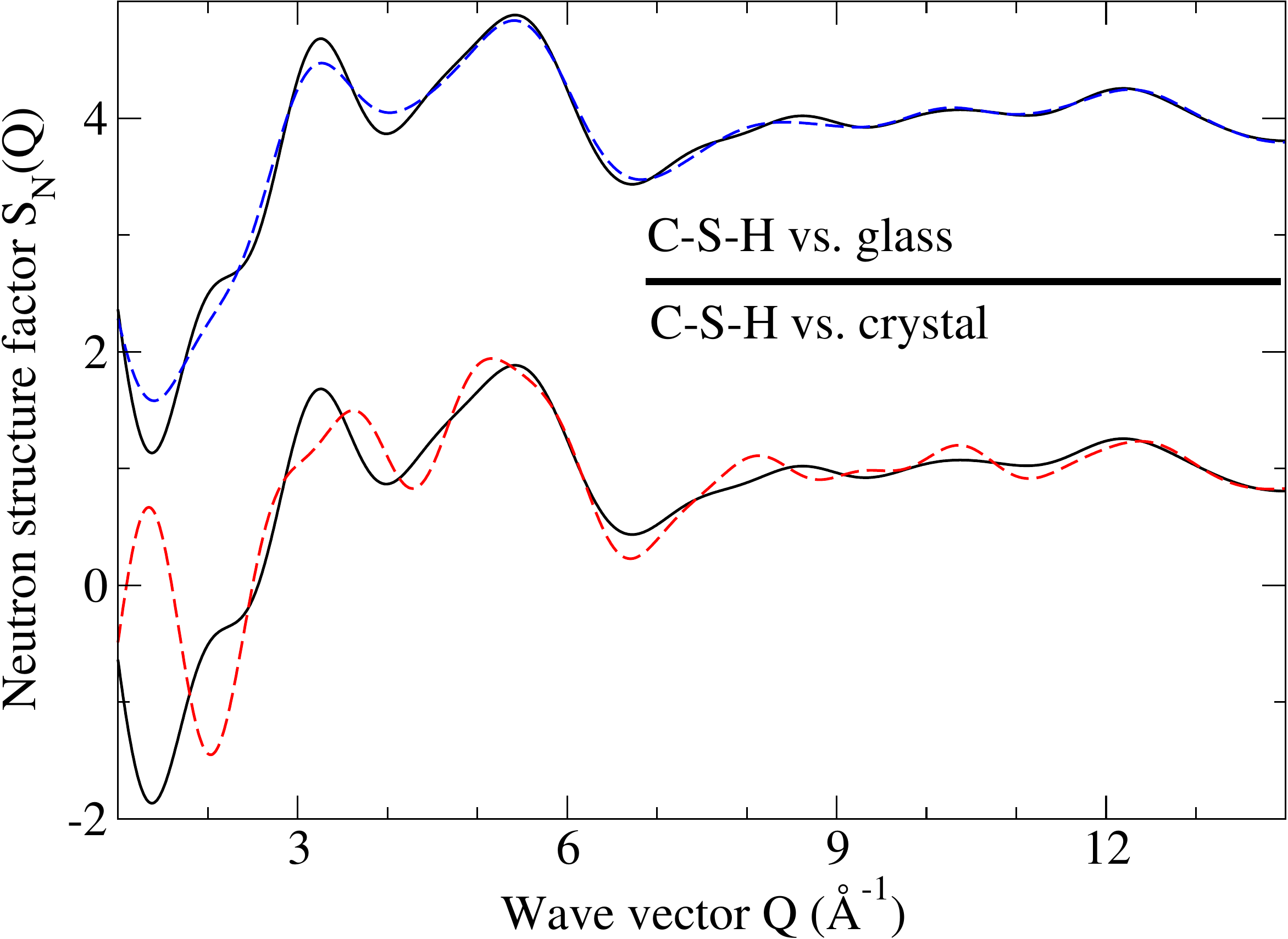}
}
\caption{Total neutron structure factor of C--S--H, compared with the one of the corresponding glass (top) and of the crystal (bottom).}
\label{fig:SN}
\end{figure}

As we realized that the local environment of the different species was not enough to distinguish the order in C--S--H from the one of the crystal and the glass, we analyzed the medium range order (MRO) of those systems by computing the neutron structure factor. The partial structure factors were first calculated from the pair distribution functions $g_{ij}(r)$:

\begin{equation}
S_{ij}(Q) = 1 + \varrho_0 \int_{0}^R 4\pi r^2 (g_{ij}(r)-1) \frac{\sin (Qr)}{Qr} F_{\text{L}}(r)\, \mathrm dr
\end{equation} where $Q$ is the scattering wave vector, $\varrho_0$ is the average atom number density and $R$ is the maximum value of the integration in real space (here $R = 6 \mathring{\text{A}}$). The $F_{\text{L}}(r) = \sin (\pi r / R) / (\pi r / R)$ term is a Lortch-type window function used to reduce the effect of the finite cutoff of $r$ in the integration \cite{wright_neutron_1988}. As discussed in the reference \cite{du_compositional_2006}, the use of this function reduces the ripples at low $Q$ but induces a broadening of the structure factor peaks. The total neutron structure factor can then be evaluated from the partial structure factors following:

\begin{equation}
S_N(Q) = (\sum_{i,j=1}^n c_ic_jb_ib_j)^{-1} \sum_{i,j=1}^n c_ic_jb_ib_j S_{ij}(Q)
\end{equation} where $c_i$ is the fraction of $i$ atoms (Si, Ca, H or O) and $b_i$ is the neutron scattering length of the species (given by 4.1491, 4.70, -3.7390 and 5.803fm for silicon, calcium, hydrogen and oxygen atoms respectively \cite{sears_neutron_1992}). Once again, note that corrections are applied to take into account the different composition and density of C--S--H with respect to tobermorite.

\subsection{Neutron structure factor}

Fig. \ref{fig:SN} shows the total neutron structure factor for C--S--H as well as for the corresponding crystal and glass. Once again, the three structure factor do not show significant differences and are typical of the ones observed in silicate \cite{micoulaut_anomalies_2013} and chalcogenide \cite{bauchy_compositional_2013} glasses. It should be mentioned that the limited sizes of the simulated systems do not allow us to study large-scale correlations (large $r$, low $Q$). Focusing on the low $Q$ region of the structure factor, we note that the peaks show the same position in C--S--H and in the glass, although they appear sharper in C--S--H, signature of an increased order on the MRO. On the contrary, the peaks at low $Q$ do not show the same positions. Since each peak is a signature of a typical spatial repetition distance in the MRO \cite{bauchy_compositional_2013}, it appears that the atomic order at intermediate length scale of C--S--H is amorphous. In particular, we observe a sharp first peak around 1.5\AA$^{-1}$ for the crystal, resulting of a strong order at large $r$, which is absent in C--S--H and the glass.

\subsection{Partial structure factors}

\begin{figure*}
\centerline{\includegraphics[height=15cm, angle=-90]{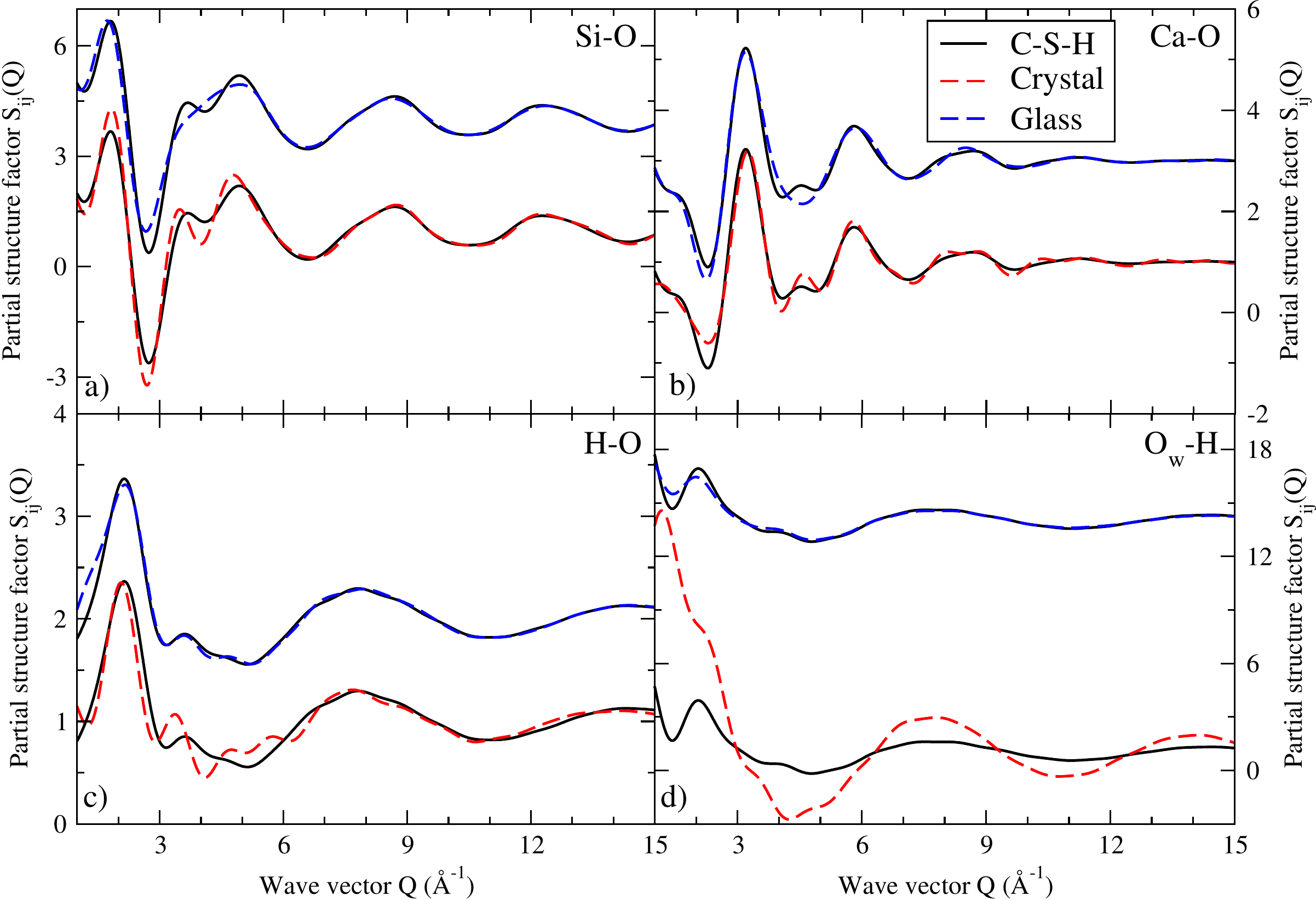}}
\caption{Partial structure factors for the Si-O (a), Ca-O (b), H-O (c), and O$_{\text{w}}$-H pairs, where O$_{\text{w}}$ refers to oxygen atoms in water molecules. For each pair, the partial structure factor of C--S--H is compared with the one of the corresponding glass (top) and of the crystal (bottom).}
\label{fig:Sij}
\centerline{\includegraphics[height=15cm, angle=-90]{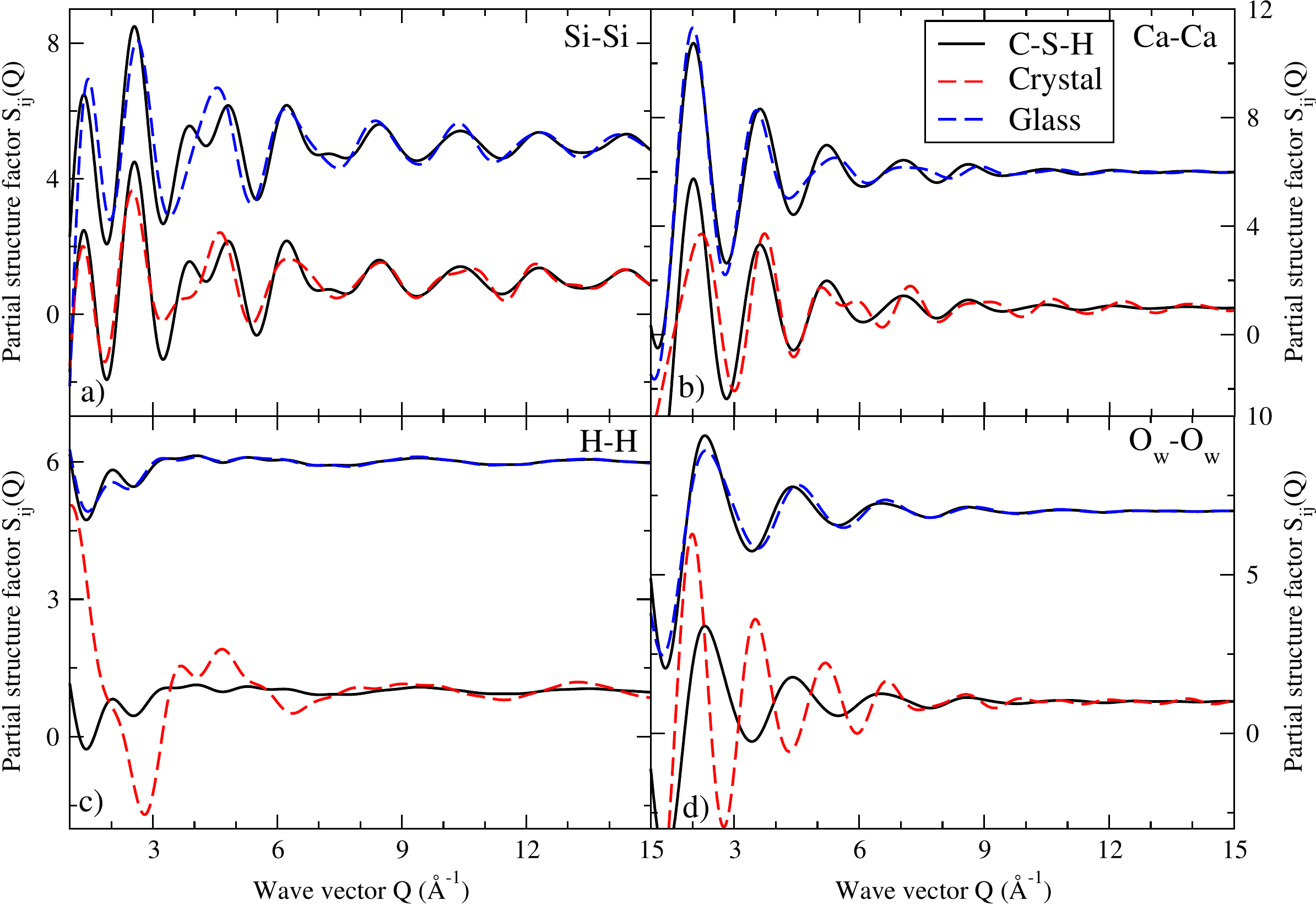}}
\caption{Partial structure factors for the Si-Si (a), Ca-Ca (b), H-H (c), and O$_{\text{w}}$-O$_{\text{w}}$ pairs, where O$_{\text{w}}$ refers to oxygen atoms in water molecules. For each pair, the partial structure factor of C--S--H is compared with the one of the corresponding glass (top) and of the crystal (bottom).}
\label{fig:Sij2}
\end{figure*}

Similarly to the case of the total PDF, the apparent disorder observed in the total structure factor actually results of some order and some disorder for different pair of species. This is what we investigated by studying the decomposition of the total structure factor into the contributions of each partial (see Fig. \ref{fig:Sij}  and \ref{fig:Sij2}). Conclusions actually appear to be the same of the ones of section \ref{sec:mro}. The MRO observed in calcium and silicate layers (tracked in the Si-O, Ca-O, Si-Si and Ca-Ca partials) for C--S--H is closer to the one of the crystal. In particular, Si-O, Ca-O and Si-Si partials respectively show peaks, respectively around 3.5, 4.5and 3.5 \AA$^{-1}$, for both C--S--H and the crystal, which are absent in the case of the glass. On the contrary, partial associated to water and hydroxyl groups are almost indentical for C--S--H and the glass, and significantly differ from those observed in the crystal. Hence, the disorder observed in the medium range order of C--S--H is mainly driven by those hydration species.

\section{Conclusion}
\label{sec:conclu}

A detailed comparison of the structure of C--S--H as well as the one of corresponding crystal and glass has allowed us to better characterize the atomic order inside the binding phase of cement. Overall, it appears that the structure of C--S--H is closer to the one of a glass than to the one of a crystal. However, some atomic order reminiscent from the one of a crystal is still found. This manifests by a fairly layered structure, by silicate chains that are not completely amorphous and by a non-random distribution of calcium atoms. On the contrary, water molecules and hydroxyl groups show a completely glassy spatial distribution, which suggests that the overall disorder is mainly caused by its hydration.

Hence, the atomic order of C--S--H can be qualified as being intermediate, definitely not fully crystalline but not perfectly amorphous. The combination of a layered structure with an overall disorder may explain the paradoxical observation of an ordered or disordered structure according to the experimental technique that is used \cite{grangeon_x-ray_2013}. Indeed, when averaged on all pairs of atom, C--S--H appears to be fully glassy. Thus, only the experiments providing an access to the detail of the order and disorder around each species would be able to capture the complexity of C--S--H.

\end{document}